\documentclass[%
 reprint,
 amsmath,amssymb,
 aps,
]{revtex4-1}

\usepackage{graphicx}
\usepackage{dcolumn}
\usepackage{bm}
\usepackage{natbib}
\usepackage[dvipsnames]{xcolor}
\usepackage{amsthm}

\begin{document}

\preprint{APS/123-QED}

\title{Quantum non-Gaussian Photon Coincidences}

\author{Lukáš Lachman}
\email{lachman@optics.upol.cz}
\author{Radim Filip}%
 \email{filip@optics.upol.cz}
\affiliation{%
 Department of Optics, Faculty of Science, Palack\' y University,\\
17. listopadu 1192/12,  771~46 Olomouc, \\
Czech Republic 
}%


\begin{abstract}
Photon coincidences represent an important resource for quantum technologies. They expose nonlinear quantum processes in matter and are essential for sources of entanglement. We derive broadly applicable criteria for quantum non-Gaussian two-photon coincidences that certify a new quality of photon sources. The criteria reject states emerging from Gaussian parametric processes, which often limit applications in quantum technologies. We also analyse the robustness of the quantum non-Gaussian coincidences and compare with the heralded quantum non-Gaussianity of single-photons based on them.
\end{abstract}

\maketitle



Five decades ago, coincidences detected in photon-counting experiments initiated the first fundamental tests of nonclassical photon pairs from nonlinear processes \cite{Clauser1974,Fry1976,Aspect1982,Friberg1985,Kuzmich2003}. These photon pairs were used to produce photon entanglement in different degrees of freedom \cite{Kwiat1999,Brendel1999}. After two decades, the effort move on to entanglement-based quantum key distribution at a significant distance and in real optical networks \cite{Xu2020}. Recently, superconducting counters of itinerate microwave photons \cite{Chen2011,Besse2018,Kurpiers2018} also allowed observation of nonclassical coincidences. This initiated the study of integrated superconducting microwave photonics beyond hybridisation in a single system \cite{Brecht2016,Brecht2017,Fan2018,Burkard2020}. Simultaneously, the correlation between optical or microwave photons and excitations in atomic \cite{Chou2005,Matsukevich2006,Moehring2007,Ritter2012}, solid-state \cite{Yilmaz2010,Usmani2012,Bernien2013}, superconducting \cite{Narla2016} and mechanical systems \cite{Riedinger2016,Riedinger2018} established a new hybrid quantum physics. A violation of Bell inequalities over a distance has already been confirmed \cite{Hensen2015}, which aims for device-independent secure key distribution \cite{Tchebotareva2019,Murta2019}.    
Therefore, photon coincidences and their analysis are crucial for the further development of many current and future experiments. 

For a long time, nonclassical photon coincidences have been in the main focus \cite{Chang2016}, as they are necessary conditions for many quantum phenomena and applications. The photon pairs produced by spontaneous parametric down-conversion, optical parametric oscillators and similar processes were the principal sources. However, the photon coincidences from such Gaussian processes still exhibit {\em multiphoton} components. This unwanted contribution grows with increasing pumping of the Gaussian process. It is known that they represent limiting factors for the rate and security of quantum key distribution \cite{Brassard2000}. Despite some solutions for particular applications, multiphoton contributions generally restrict the speed and performance of any entanglement-based photonic protocols, which already achieve a considerable distance \cite{Yin2017}. The states with reduced multiphoton contributions will expand current photonic quantum technology.
Currently developing experimental platforms brought new versions of two-photon optical processes in atoms \cite{Wilk2007,Hacker2016,Hamsen2017} and solid-state systems \cite{Benson2000,Fattal2004,Stevenson2006,Akopian2006,Munoz2014} with rapid advances \cite{Jayakumar2013,Jayakumar2014,Huber2018,Reindl2018,Chen2018,Prilmueller2018,Basset2019,Wang2019,Liu2019}, but also with single-atom mechanical oscillators \cite{Ding2017,Ding2018,Gasparinetti2017} and at microwave frequencies \cite{Gasparinetti2017} and soon in other superconducting circuits \cite{Frattini2018}. This effort even extends to observable three-photon coincidences \cite{Khoshnegar2017}. They are all capable of producing photon pairs with much lower multiphoton contributions than Gaussian processes. 

These experimental developments allow us to test, for the first time, that they conclusively reach the capability to produce {\em quantum non-Gaussian coincidences} that are better than coincidences provided by any correlated Gaussian states of light. The eminent first target is a coincidence of Fock states $|1\rangle|1\rangle$ in two different modes without any higher photon contribution. As described above, it is ideal for building two-photon entanglement without multiphoton components, but also other applications in multiplexing of Fock-state-based quantum sensing \cite{Wolf2019}.  In this Letter, we derive ab initio criteria for quantum non-Gaussian coincidences for commonly used experimental setups with multiplexed single-photon detectors, analyse their essential features and robustness and modify the method for currently developing photon number detectors. Remarkably, low photon rates do not preclude observation of quantum non-Gaussian coincidences. Therefore they are applicable to the majority of above mentioned experimental platforms even at an early stage of their development. This result qualitatively extends the already experimentally verified quantum non-Gaussian statistics of heralded single-mode states \cite{Straka2014}. The quantum non-Gaussianity was also recognized in light from quantum dots \cite{Predojevic2014}, which have the potential to test quantum non-Gaussian coincidences. Therefore, we can compare these two different quantum non-Gaussian statistics of photons. Since nonclassicality is a necessary condition of the quantum non-Gaussianity, we analyse both quantum non-Gaussian statistics of photons in the experimental layout where nonclassical coincidences are detected \cite{Jayakumar2013} to understand the difference from the standard nonclassical tests.

\begin{figure}[t]
\centerline {\includegraphics[width=0.95\linewidth]{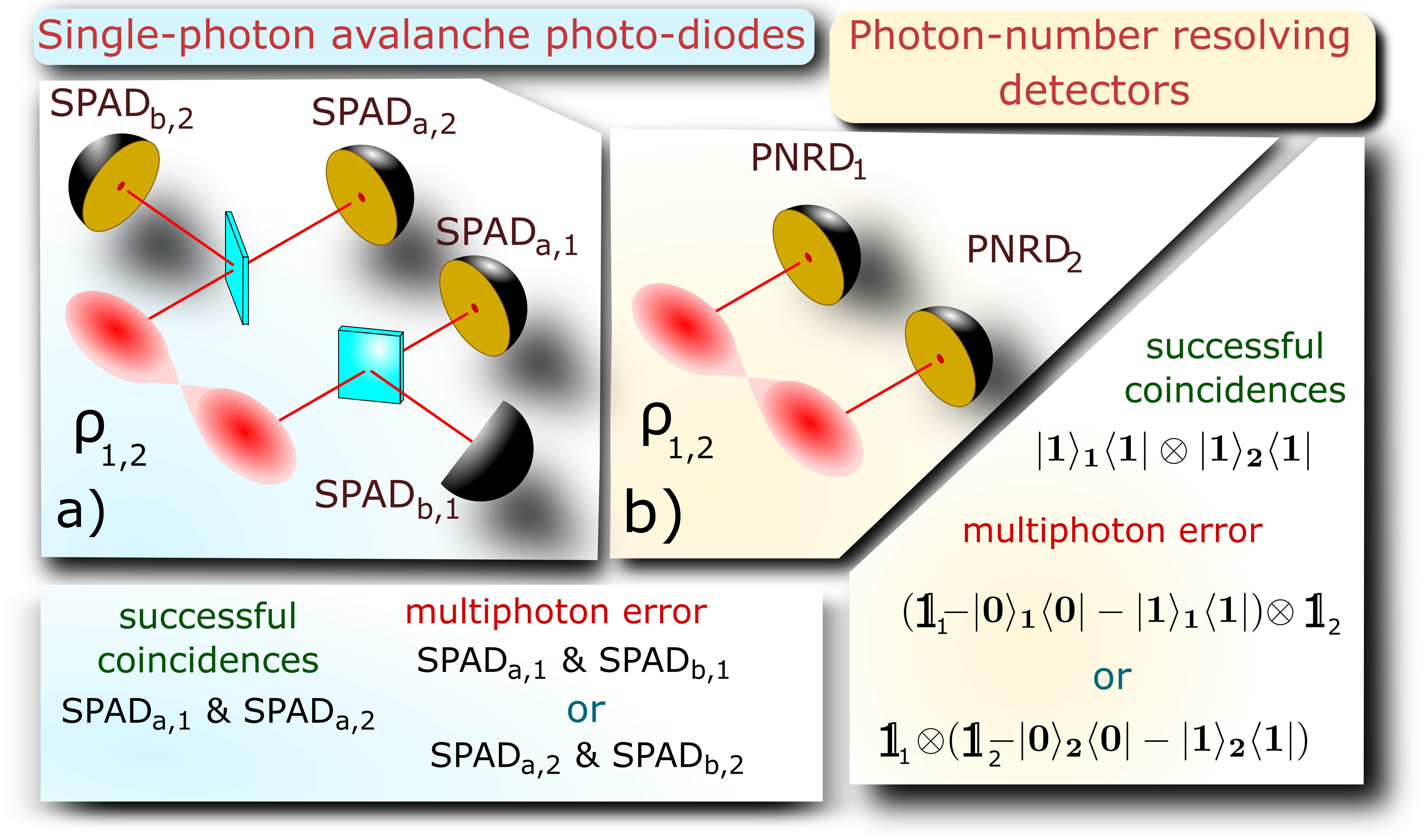}}
\caption{Schemes enabling detection of the nonclassical and quantum non-Gaussian coincidences using single-photon avalanche photo-diode ({\emph a}) and photon-number resolving detector ({\emph b}). In the figure ({\emph a}), two optical modes are split by a balanced BS and measured by two pairs of single-photon avalanche photo-diodes. The panel below shows which detector clicks are important for disclosing both quantum aspects. In the figure ({\emph b}), both modes are measured directly by two photon-number resolving detectors. They discriminate the single-photon income from two and more-photons income. The right lower panel summarizes the employed positive valued operators giving the successful coincidences and multi-photon error.}
\label{fig:scheme}
\end{figure}

{\em Nonclassical coincidences---} A measurement result rejecting interpretation of detected radiation as the classical waves signifies the nonclassicality. If coincidences of detection events from sensitive single-photon photodetectors in two modes of radiation cannot be obtained by classical waves, we denote such coincidences as nonclassical. Specifically, the nonclassical coincidences manifest themselves in a layout using single-photon avalanche diodes (SPADs) that is depicted in Fig.~\ref{fig:scheme} a). Light in two distinguishable modes denoted as $1$ and $2$ propagates through a beam-splitter (BS) in each mode and two pairs of SPADs measure the split light. A criterion compares probabilities of simultaneous clicks at selected pairs of detectors as summarized in the panel below. A success probability $P_s$ quantifies events when the detectors SPAD$_{a,1}$ and SPAD$_{a,2}$ click simultaneously. The error probability $P_{e,i}$  with $i=1,2$ measures when the detectors SPAD$_{a,i}$ and SPAD$_{b,i}$ click.  Thus, the success events represent coincidences occurring in different modes and error events correspond to coincidences in the same input mode. The criterion stems from a linear combination of those probabilities \cite{Filip2011}
\begin{equation}
F_{a}(\rho)=P_s+a ( P_{e,1}+ P_{e,2}),
\label{FaRho}
\end{equation}
where $a$ is a free parameter. To achieve the nonclassicality criterion for such measurement, the functional $F_{a}(\rho)$ is maximized over all states of classical waves
\begin{equation}
\rho \neq \int P(\alpha_1,\alpha_2)\vert \alpha_1 \rangle_1 \langle \alpha_1 \vert \otimes \vert \alpha_2 \rangle_2 \langle \alpha_2 \vert \mathrm{d}^2 \alpha_1 \mathrm{d}^2 \alpha_2,
\label{clStates}
\end{equation}
where $P(\alpha,\beta)$ is a density probability function and subscripts $1$ and $2$ distinguish the two modes. The optimizing leads to a threshold function $F(a)=0$ for $a\leq -1/2$ and $F(a)=2a+1$ for $a>-1/2$, which covers all outputs of classical states. The nonclassicality happens when $P_s>\min_a F(a)-a P_e$ \cite{Filip_2013}. Exclusion of the parameter $a$ induces the nonclassicality criterion
\begin{equation}
\frac{2 P_s}{P_{e,1}+P_{e,2}}>1
\label{probCrit1}
\end{equation}
Simultaneous generation of pairs $|1,1\rangle$ of single-photon states without any multiphoton contributions is always detected as nonclassical.
 Note, the commonly used criterion $P_s^2/(P_{e,1}P_{e,2}) \leq 1$ \cite{Kuzmich2003} from the Cauchy-Schwarz inequality reveals the nonclassicality identically for realistic states producing the error events symmetrically, i. e. $P_{e,1}=P_{e,2}$. Besides that, the condition (\ref{probCrit1}) gets stricter in general. To continue tests of fundamental aspects of the photon pairs and its generation, the threshold needs to be moved up to reject all two-mode Gaussian states from parametric processes governed by the quadratic interaction Hamiltonians.
 


\begin{figure}[t]
\centerline {\includegraphics[width=0.9\linewidth]{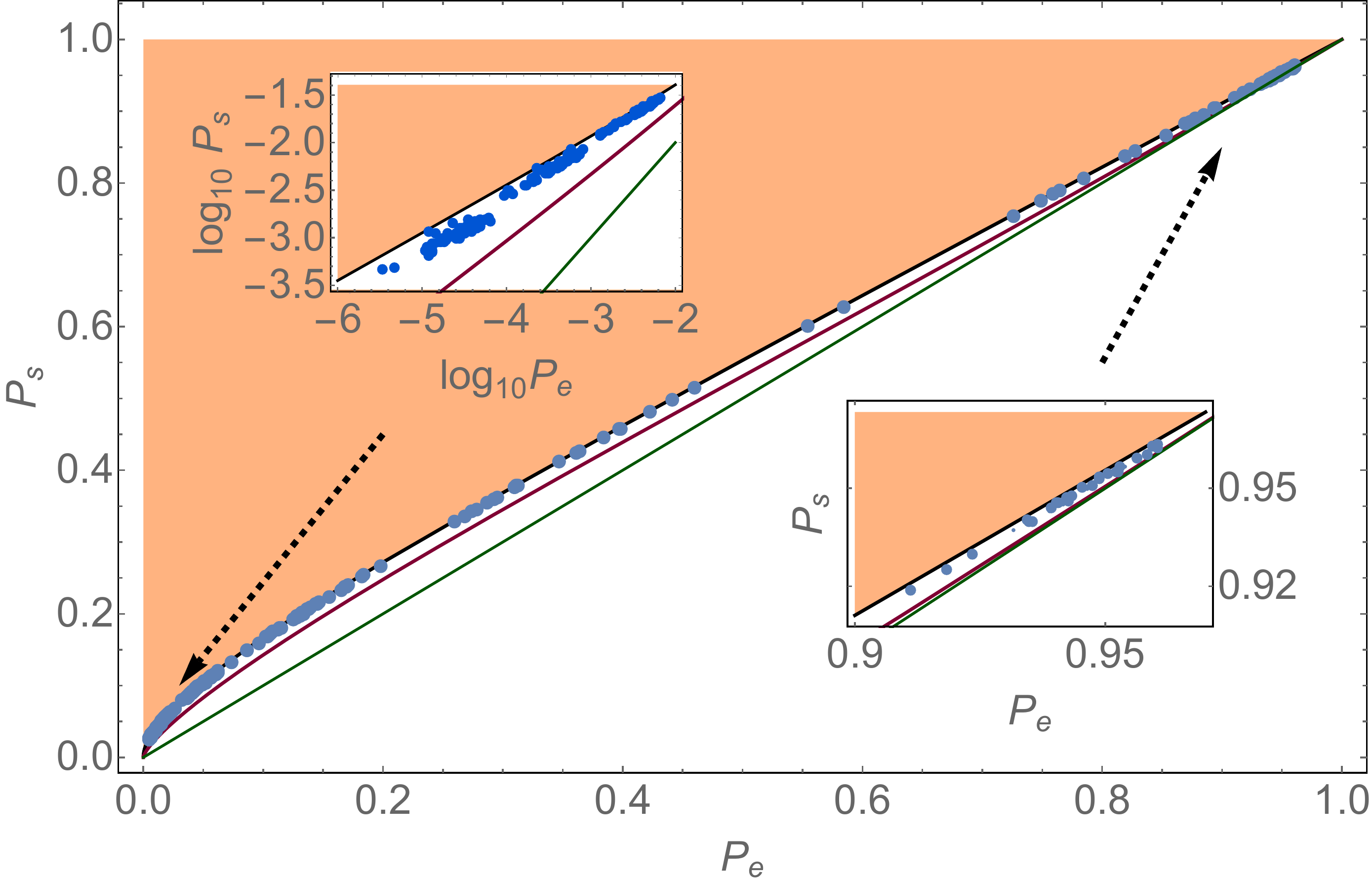}}
\caption{The quantum non-Gaussian coincidences are recognized when the measured probabilities $P_s$ and $P_{e}$ belong to the orange region. The solid black line corresponds to the threshold determined by relation (\ref{thresSym}). Its reliability was verified by a Monte-Carlo simulation producing in total $10^7$ random Gaussian states, see the Supplemental Material for more details. Five hundred best attempts are depicted by the blue points. The purple line shows a threshold for two factorizable Gaussian states to visualize that states with Gaussian photon correlation can be above them. The green line corresponds to the threshold of nonclassicality (\ref{probCrit1}). The upper inset presents the thresholds and results of the Monte-Carlo simulation in a region of very attenuated states which is relevant for many experiments \cite{Rota2020}. The lower inset zooms the results in the corner with very high probabilities of success and error.}
\label{fig:mc}
\end{figure}

{\em Quantum non-Gaussian photon coincidences---} Quantum non-Gaussianity denotes states going beyond mixtures of Gaussian states. Let us focus for first time on the recognition of the quantum non-Gaussian states that occupy two modes. Formally, quantum non-Gaussianity in two modes is defined as
\begin{equation}
\rho \neq \int P\left( G \right) \vert G \rangle_{1,2} \langle G \vert \mathrm{d}^2G,
\label{QNG}
\end{equation}
where $\vert G \rangle_{1,2} $ is a two-mode Gaussian state and $P(G)$ is a density probability function of parameters identifying the state $\vert G \rangle_{1,2}$. The schemes at Fig.~\ref{fig:scheme} allows us to distinguish those quantum non-Gaussian states by passing a criterion for quantum non-Gaussian coincidences, which rejects even states exhibiting Gaussian photon correlations, which are typical for linearized dynamics from quadratic nonlinearities \cite{Harder2016}. Firstly, we inspect the scheme in Fig.~\ref{fig:scheme} a), where SPADs are exploited for the detection. As in the case of nonclassicality, optimizing the functional
\begin{equation}
    F_a(\rho)=P_s+a (P_{e,1}+P_{e,2})
    \label{FG}
\end{equation}
induces the criterion. Because the functional is linear, the optimum is given by a pure two-mode Gaussian state  \cite{Filip2011}. We use the Bloch Messiah reduction \cite{Braunstein2005} to parametrize all these pure Gaussian states.
To establish a criterion, we have to optimize the linear combination (\ref{FG}) over eight parameters determining all the states $\vert G \rangle_{1,2}$. The Supplemental Material instructs how to derive analytical but extensive formulas for the probabilities $P_s$ and $P_e=(P_{e,1}+P_{e,2})/2$. To perform the maximizing, we certify a conjecture that some two-mode squeezed state $\vert G_{r} \rangle_{1,2} = \sqrt{1- r^2}\sum_{n=0}^{\infty} r^n \vert n\rangle_1 \vert n \rangle_2$ maximizes (\ref{FG}) for a given $a$. Under the assumption, the derived threshold function $F(a)$ induces a condition
\begin{equation}
    P_s>\frac{1}{2}\sqrt{\frac{P_e}{8+P_e}}\left[2 +P_e+\sqrt{P_e(8+P_e)}\right].
    \label{thresSym}
\end{equation}
Analytical proof that $\vert G_{r} \rangle_{1,2}$ yields the global maximum of (\ref{FG}) is too challenging. Thus, we chose two different approaches to deal with that. First, we performed a Monte-Carlo simulation where the Gaussian states were randomly generated to certify the threshold (\ref{thresSym}). Second, we considered the function (\ref{FG}) with $-a \gg 1$ for which the optimal Gaussian states obey experimentally typical case of $P_e \ll 1$. The Supplemental Material provides a proof that the state $\vert G_r \rangle$ represents the global optimum in this experimentally relevant limit.
In Fig.~\ref{fig:mc}, we compare criterion (\ref{probCrit1}) in green with the criterion for quantum non-Gaussian coincidences (\ref{thresSym}) in black. The figure also compares these thresholds with a purple line covering all mixtures of factorized Gaussian states. It highlights that states with Gaussian correlations establish the condition on quantum non-Gaussian coincidences. The demands of the criteria (\ref{probCrit1}) and (\ref{thresSym}) will be analyzed later on a particular model of experimentally relevant states.


{\em Quantum non-Gaussian coincidences for detection with PNRDs---}
Modern detection technique employs photon-number resolving detectors (PNRDs) instead of SPADs \cite{Harder2016}. The layout in Fig.~\ref{fig:scheme} b) modifies the experimental scheme for this situation. A PNRD allows us to distinguish the number of arriving photons. Two PNRDs responding on different modes quantify the probability $P_{m,n}=\langle m \vert \langle n \vert \rho \vert n \rangle \vert m \rangle$. Following the approach, we define the success probability by $P_s=P_{1,1}$. The error probability $P_{e,i}$ corresponds to probability of multiphoton contributions in the  $i$th mode, i. e. $P_{e,i}=1-P_{0}^{(i)}-P_{1}^{(i)}$, where $P_{n}^{(i)}$ is the photon number distribution in the $i$th mode with the other mode being ignored. The state $\vert G_r \rangle$ establishes a criterion of quantum non-Gaussian coincidences in the form
\begin{equation}
P_{s} >\sqrt{P_{e}}-P_{e}
\label{PNRDThres}
\end{equation}
in this detection scheme, where $P_e$ stands for the average of error probabilities again. The covering of all mixtures of Gaussian states was verified by a Monte-Carlo simulation as well. Thus, the criterion (\ref{PNRDThres}) can be used specifically for two-quanta experiments where two-mode photon number statistics is detectable using optical homodyne tomography \cite{Makino2016}, in microwave experiments \cite{Gao2018} and trapped ions experiments \cite{Ding2017,Ding2018}.


{\em Testing experimental example---} Applicability of the criteria for quantum non-Gaussian coincidences can be illustrated on an example of a model state that is relevant for modern quantum technologies with atoms or solid state emitter in the two-mode cavities \cite{Gines2021}. They exploit a cascade energy transfer in matter to radiate a correlated pair of photons with a density matrix approaching \cite{Akopian2006,Liu2019}
\begin{equation}
\rho_{1,2}(\eta)=\eta \vert 1 \rangle_1 \langle 1 \vert \otimes \vert 1 \rangle_2 \langle 1 \vert+(1-\eta)\vert 0 \rangle_1 \langle 0 \vert \otimes \vert 0\rangle_2 \langle 0 \vert,
\label{modelSig}
\end{equation}
where $\eta$ is the probability that a photon pair $\vert 1, 1 \rangle$ is generated. 
However, such a source typically suffers from high losses and noise deteriorating the photon statistics and a density matrix of the radiated light obtains
\begin{equation}
\begin{aligned}
    \rho &=\mbox{Tr}_{3,4}\{ L_{2,4}(T_2)L_{1,3}(T_1) \cdot \left[\mathcal{N}_{\bar{n}_1,\bar{n}_2}(\rho_{1,2}) \otimes \vert 0 \rangle_3 \langle 0 \vert \right. \\
    &\left. \otimes \vert 0 \rangle_4 \langle 0 \vert \right] \cdot L_{2,4}^{\dagger}(T_2)L_{1,3}^{\dagger}(T_1)\},
\end{aligned}
\label{model}
\end{equation}
where $L_{i,j}(T)$ corresponds to unitary operator characterizing the optical loss from the mode $i$ to the mode $j$. Tracing the modes $3$ and $4$ gives rise to a state affected by losses in modes $1$ and $2$ with the transmission $T_1$ and $T_2$. The trace-preserving map  $\mathcal{N}_{\bar{n}_1,\bar{n}_2}$ add the noise to the both modes. The parameter $\bar{n}_i$ quantifies the mean number of noisy photons in the $i$th mode.
Both the losses and the noise reduce the coincidences, i. e. the component $\vert 1, 1 \rangle$.

Considering experimentally relevant weak emission with strongly suppressed multiphoton contributions, the success probability is $P_s \approx T_1 T_2 \eta\left[1+\bar{n}_1(1-T_1)+\bar{n}_2(1-T_2)\right]/4 +T_1 T_2 \bar{n}_1 \bar{n}_2/4$ and the error probabilities approach $P_{e,i} \approx \eta T_i^2 \bar{n}_i+T_i^2\bar{n_i}^2/4$,
where we assume $\bar{n}_i \ll 1$ without any conjecture about losses and the parameter $\eta$.
According to them, the considered state exhibits the nonclassicality if $\eta \gtrapprox (T_1 \bar{n}_1-T_2 \bar{n}_2)^2/(2T_1 T_2)$. 
In contrast, the quantum non-Gaussian coincidences are observed only for much better sources emitting the states modeled by (\ref{model}). Analytical conditions on the state are derived only for the considered limit. Employing the relation approximating the threshold $P_s^2 \approx P_{e}/8$, gives rise to a condition
\begin{equation}
\begin{aligned}
\eta & \gtrapprox \frac{1}{2 T_1^2 T_2^2}\left[\bar{n}_1 T_1^2+\bar{n}_2 T_2^2+\right.\\
& \left. 4\sqrt{(\bar{n}_1 T_1^2-\bar{n}_2 T_2^2)^2-T_1^2T_2^2 (\bar{n}_1^2 T_1^2-\bar{n}_2^2 T_2^2)}\right].
\end{aligned}
\label{condNonG1}
\end{equation}
It shows how this quantum aspect is sensitive to the noise contributions in this regime. Assuming $T_1=T_2=T \ll 1$ in formula (\ref{condNonG1}) allows us to estimate the depth  of quantum non-Gaussian coincidences \cite{Straka2014}
\begin{equation}
T \approx \left[\frac{\bar{n}_1+\bar{n}_2}{\eta}\right]^{1/2},
\label{condNonG2}
\end{equation}
 Note, the model state (\ref{model}) occupies two modes, and therefore the criteria (\ref{thresSym}) and (\ref{PNRDThres}) can be used for the evaluation of the quantum non-Gaussian coincidences. The Supplemental Material includes proposed evaluation of states occupying many modes together with an accurate analysis of when the state in (\ref{model}) manifests the quantum non-Gaussian coincidences.
 
\begin{figure}[t]
\centerline {\includegraphics[width=0.9\linewidth]{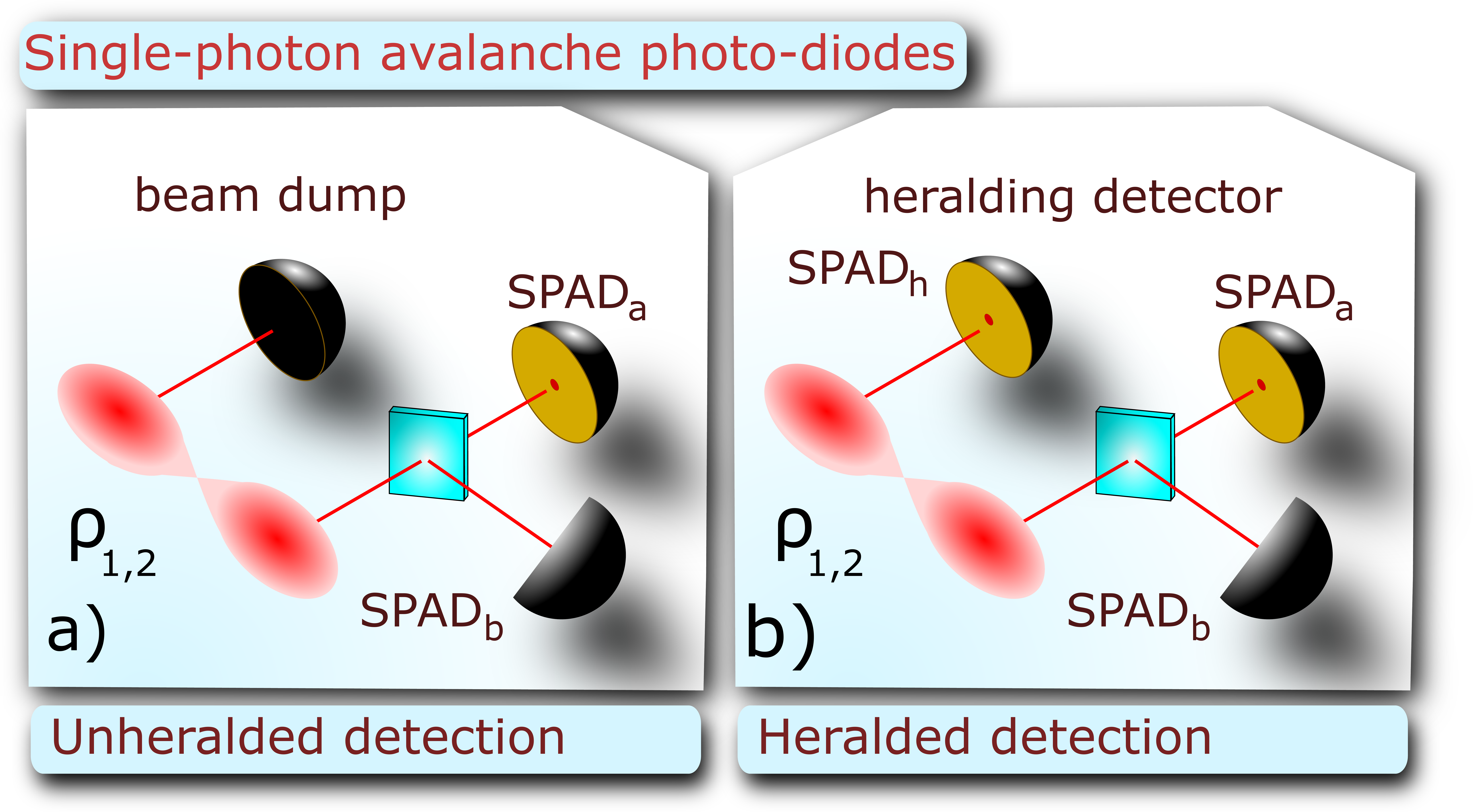}}
\caption{The single-mode quantum non-Gaussianity from two-mode states: ({\emph a}) The measurement is performed by splitting light in one mode towards two SPADs while the second mode is dumped (ignored) \cite{Straka2014}. ({\emph b}) The dumped mode in case ({\emph a}) is now used for heralding that selects the detection events in SPAD$_a$ and SPAD$_b$ according to a respond SPAD$_{\mbox{h}}$.}
\label{fig:scheme2}
\end{figure}

{\em Heralded and unheralded single-mode quantum non-Gaussianity---} The considered model allows us to compare the quantum non-Gaussian coincidences with both heralded and unheralded detection of the quantum non-Gaussianity in a single-mode, which was explored theoretically \cite{Filip2011,Lachman2013} and measured \cite{Jezek2011, Straka2014}. In both cases depicted in Fig.~\ref{fig:scheme2}, the measurement of the single-mode quantum non-Gaussianity is performed by a beam-splitter that divides photons towards two detectors SPAD$_a$ and SPAD$_b$ responding only on one of the modes, which the inspected light occupies. The other mode is damped (\emph{a}) or used for heralding (\emph{b}). A criterion for the outputs of this detection scheme was derived in Ref. \cite{Lachman2013}. The density matrix of the state (\ref{model}) in the measured mode obtains formally the same form for both the schemes that is given by
\begin{equation}
    \rho=\mbox{Tr}_2\left \{ L_{1,2}(T)\cdot \left[\mathcal{N}_{\bar{n}}(\rho_{\eta_s}) \otimes \vert 0 \rangle_2 \langle 0 \vert \right]\cdot L_{1,2}^{\dagger}(T)\right \},
    \label{reduced}
\end{equation}
where $\rho_{\eta_s}=\eta_s \vert 1 \rangle_1 \langle 1 \vert+(1-\eta_s)\vert 0 \rangle_1 \langle 0 \vert$ and $L_{1,2}(T)$ and $\mathcal{N}_{\bar{n}}$ are defined identically as in (\ref{model}). The parameter $\eta_s$ obtains $\eta_{s}=\eta$ for the unheralded scheme. For the scheme with heralding, $\eta_{s}$ yields
\begin{equation}
    \eta_{s}=\eta T_h \frac{1-e^{-\bar{n}_h T_h}(1-T_h+\bar{n}_h T_h^2)}{1-e^{-\bar{n}_h T_h}\left(1-\eta T_h+\eta\bar{n}_h T_h^2\right)},
\end{equation}
where $T_h$ and $\bar{n}_h$ quantify the losses and the noise contributions in the heralding mode.
 For the states with low noise contributions $\bar{n} \ll 1$, the approximate criterion obtains the form $P_s^3>P_e/4$ where $P_s$ denotes a probability of a click occurring on SPAD$_{a}$ and $P_e$ stands for simultaneous clicks of both detectors SPAD$_{a}$ and SPAD$_{b}$ in Figs.~\ref{fig:scheme2}. The test of quantum non-Gaussianity requires \cite{Lachman2013}
\begin{equation}
    \eta > \sqrt{2\bar{n}/T}
    \label{condNonG3}
\end{equation}
 for the unheralded case. Comparing relations (\ref{condNonG1}) and (\ref{condNonG2}) shows that quantum non-Gaussian photon coincidences survive lower photon-pair emission $\eta$ but they are more sensitive to losses than unheralded single-mode quantum non-Gaussianity \cite{Higginbottom2016}. When heralding is used for a state preparation, the quantum non-Gaussianity of heralded states manifests itself when $T > 2\bar{n}$ regardless of the parameter $\eta$ \cite{Straka2014}. Thus, the single-mode quantum non-Gaussianity is revealed more easily with the help of heralding than the quantum non-Gaussian coincidences. However, it gives no evidence about the quantum non-Gaussianity of the unheralded states because the heralding can prepare single-mode quantum non-Gaussianity from the Gaussian states \cite{Jezek2011,Straka2014}.
 

{\em Conclusion and outlook---} We extended quantum non-Gaussianity of single mode states \cite{Straka2018,Lachman2019}  to quantum non-Gaussian coincidences between two modes of light, microwaves or phonons of mechanical oscillators. The proposed methods are directly applicable to the two-mode versions of optical experiments with atomic systems \cite{Wilk2007,Hacker2016}, two-photon solid-state emitters \cite{Jayakumar2014,Huber2018,Reindl2018,Chen2018,Prilmueller2018,Basset2019}, but also to upcoming electromechanical experiments \cite{Chu2018,Sletten2019}, quantum mechanics with trapped ions \cite{Ding2017,Ding2018} and two-mode superconducting circuits \cite{Gao2018,Gao2019}. A straightforward theoretical extension is evaluation of the multiphoton quantum non-Gaussian coincidences of Fock states $\vert n \rangle \vert m \rangle$ to investigate multiphoton and multiphonon nonlinear process. It can be applied to time-bin experiments with single-photon guns to test prepared coincidences \cite{Kuhn2002,Lodahl2004,Peter2005,Chu2016}. Simultaneously, the approach can be extended to exposing the quantum non-Gaussianity of sources producing triplets of photons \cite{Khoshnegar2017}.



\begin{acknowledgments}
We thank Jarom\' ir Fiur\' a\v sek for a fruitful discussion. We acknowledge the support from the Czech Science Foundation under the project 20-16577S. This work has received national funding from the MEYS and the funding from European Union’s Horizon 2020 (2014-2020) research and innovation framework programme under grant agreement No 731473 (project 8C18002). Project HYPER-U-P-S has received funding from the QuantERA ERA-NET Cofund in Quantum Technologies implemented within the European Union’s Horizon 2020 Programme. L. L. acknowledges internal projects of Palack\' y University IGA-PrF-2020-009. R.F. acknowledges project LTAUSA19099 from the Ministry of Education, Youth and Sports of Czech Republic. 
\end{acknowledgments}

\bibliography{coincidences.bib}
\vspace{1cm}
\centerline {{\textbf{\large Supplemental Materials}}}

\section{Gaussian states}
Unitary transformations of Gaussian states occupying $n$ modes are conveniently described by transformation of the covariance matrix $\boldsymbol{\Gamma}$ and the vector of the first moments $\boldsymbol{V}$. The covariance matrix has elements
\begin{eqnarray}
\Gamma_{2i-1,2j-1}&=&\frac{1}{2}(\langle X_i X_j \rangle+\langle X_j X_i \rangle) -\langle X_i \rangle \langle X_j \rangle \nonumber \\
\Gamma_{2i,2j}&=&\frac{1}{2}(\langle P_i P_j \rangle+\langle P_j P_i \rangle) -\langle P_i \rangle \langle P_j \rangle \nonumber \\
\Gamma_{2i-1,2j}&=&\frac{1}{2}(\langle X_i P_j \rangle+\langle P_j X_i \rangle) -\langle X_i \rangle \langle P_j \rangle \nonumber \\
\Gamma_{2i,2j-1}&=&\frac{1}{2}(\langle P_i X_j \rangle+\langle X_j P_i \rangle) -\langle P_i \rangle \langle X_j \rangle,
\label{covMat}
\end{eqnarray}
where $i,j$ index the considered modes, $X_i$ is the coordinate operator and $P_i$ is the momentum operator. The vector $\boldsymbol{V}$ has elements
\begin{eqnarray}
V_{2i-1}&=&\langle X_i \rangle \nonumber \\
V_{2i}&=& \langle P_i \rangle.
\end{eqnarray}
The covariance matrix together with the vector $\boldsymbol{V}$ specify any Gaussian state.

 The unitary operations preserving the Gaussian states are squeezing, rotation of the coordinates corresponding to the free evolution and the beam splitter transformation. All these operations can be represented by matrices that transforms $\boldsymbol{\Gamma}$ and $\boldsymbol{V}$. To complete the Gaussian transformations, we also introduce the displacement operator $\boldsymbol{D}(\alpha)$ that is represented by a vector. It affects only the vector $\boldsymbol{V}$. According to the Bloch-Messiah reduction \cite{Braunstein2005}, any Gaussian state occupying $N$ modes can be prepared by squeezing every mode, followed by mixing the modes on beam splitters and, finally, acting the displacement operators on the emerging states.
 
 Let us provide all these unitary operators in this notation. Let the matrix $\boldsymbol{S}^{(i)}$ represents squeezing acting only the $i$th mode.  For $\xi$ being real, the $\boldsymbol{S}^{(i)}$ has elements
\begin{eqnarray}
S_{2i-1,2i-1}^{(i)}&=& \exp(-\xi) \nonumber \\
S_{2i,2i}^{(i)}&=& \exp(\xi)
\end{eqnarray}
and $S_{m,n}^{(i)}=\delta_{m,n}$ otherwise. The rotation matrix $\boldsymbol{R}^{(i)}(\phi)$ acting on the $i$th mode is given by
\begin{eqnarray}
R_{2i-1,2i-1}^{(i)}&=&R_{2i,2i}=\cos \phi \nonumber \\
R_{2i-1,2i}^{(i)}&=&-R_{2i,2i-1}=\sin \phi
\end{eqnarray}
and $R_{m,n}^{(i)}=\delta_{m,n}$ otherwise. A general squeezing operator $\boldsymbol{S}(\xi)$ affecting the $i$th mode obtains
\begin{equation}
\boldsymbol{S}^{(i)}(\xi)=\boldsymbol{R}^{(i)}(-2\phi)\boldsymbol{S}^{(i)}(\vert \xi \vert) \boldsymbol{R}^{(i)}(2\phi),
\end{equation}
where $\xi=\vert \xi \vert e^{i \phi}$. The matrix $\boldsymbol{U}_{BS}^{(i,j)}(\tau)$ corresponding to a beam splitter that transforms the modes $i$ and $j$ is
\begin{eqnarray}
U_{BS,2i-1,2i-1}^{(i,j)}&=&U_{BS,2i,2i}^{(i,j)}=\sqrt{\tau} \nonumber \\
U_{BS,2j-1,2j-1}^{(i,j)}&=&U_{BS,2j,2j}^{(i,j)}=\sqrt{\tau} \nonumber \\
U_{BS,2i-1,2j-1}^{(i,j)}&=&U_{BS,2i,2j}^{(i,j)}=\sqrt{1-\tau} \nonumber \\
U_{BS,2j-1,2i-1}^{(i,j)}&=&U_{BS,2j,2i}^{(i,j)}=-\sqrt{1-\tau}
\end{eqnarray}
and $U_{BS,m,n}^{(i,j)}=\delta_{m,n}$ otherwise. The displacement is represented formally by a vector 
\begin{eqnarray}
\boldsymbol{D}(\alpha)&=&(\vert \alpha_1 \vert \cos \psi_1,\vert \alpha_1 \vert \sin \psi_1, \nonumber \\
&\ &...,\vert \alpha_n \vert \cos \psi_n,\vert \alpha_n \vert \sin \psi_n)
\end{eqnarray}
which carries out transformation
\begin{equation}
\boldsymbol{V}=\widetilde{\boldsymbol{V}}+\boldsymbol{\Gamma}\boldsymbol{D},
\end{equation}
where $\widetilde{\boldsymbol{V}}$ is the vector of the first moments before an action of the displacement operator. The covariance matrix remains the same under this transformation.

The covariance matrix of a general state $\vert G \rangle$ propagating through the setup in Fig.~1 a) of the main text is determined by \cite{Braunstein2005}
\begin{eqnarray}
&\ &\boldsymbol{\Gamma}=\boldsymbol{U}_{BS}^{(1,2)}(1/2)\boldsymbol{U}_{BS}^{(3,4)}(1/2)\boldsymbol{U}_{BS}^{(2,3)}(\tau) \boldsymbol{S}^{(2)}(\xi_2)\boldsymbol{S}^{(1)}(\xi_1) \cdot\nonumber \\
&\ & \mathbb{I} \cdot \boldsymbol{S}^{(1),T}(\xi_1) \boldsymbol{S}^{(2),T}(\xi_2)\boldsymbol{U}_{BS}^{(2,3),T}(\tau)\nonumber \\
&\ &\boldsymbol{U}_{BS}^{(3,4),T}(1/2)\boldsymbol{U}_{BS}^{(1,2),T}(1/2),
\end{eqnarray}
where the superscript $T$ denotes the transposition of the matrix. The first moments yield
\begin{equation}
\boldsymbol{V}^T=\boldsymbol{U}_{BS}^{(1,2)}(1/2)\boldsymbol{U}_{BS}^{(3,4)}(1/2) \widetilde{\boldsymbol{\Gamma}} \boldsymbol{D}^T(\alpha)
\end{equation}
with
\begin{eqnarray}
\widetilde{\boldsymbol{\Gamma}}&=&\boldsymbol{U}_{BS}^{(2,3)}(\tau) \boldsymbol{S}^{(2)}(\xi_2)\boldsymbol{S}^{(1)}(\xi_1) \cdot\nonumber \\
&\ & \mathbb{I} \cdot \boldsymbol{S}^{(1),T}(\xi_1) \boldsymbol{S}^{(2),T}(\xi_2)\boldsymbol{U}_{BS}^{(2,3),T}(\tau)
\end{eqnarray}
and
\begin{eqnarray}
\boldsymbol{D} &=& (\vert \alpha_1 \vert \cos \psi_1,\vert \alpha_1 \vert \sin \psi_1, \nonumber \\
&\ &\vert \alpha_2 \vert \cos \psi_2,\vert \alpha_2 \vert \sin \psi_2,0,0,0,0).
\end{eqnarray}
Projection on the vacuum in one or more modes is given by \cite{Weedbrook2012}
\begin{equation}
P_{\boldsymbol{M}}=\frac{\exp \left[ \frac{\boldsymbol{V}(\boldsymbol{\Gamma}+\boldsymbol{M})^{-1}\boldsymbol{V}^T-\boldsymbol{V}\boldsymbol{\Gamma} \boldsymbol{V}^T}{2}\right]}{\sqrt{\det (\boldsymbol{\Gamma}+\boldsymbol{M})}}
\label{p0G}
\end{equation}
with $\boldsymbol{M}$ being a matrix determining the measurement with elements $M_{i,j}=\delta_{i,j}m_i$, where  $m_{2 k-1}=m_{2k}=1$ if the projection is carried out in the $k$th mode and otherwise $m_{2 k-1}=m_{2k}=0$. For a simpler notation, let us introduce a vector $\boldsymbol{m}=(m_1,...,m_{2n})$ and distinguish the probabilities (\ref{p0G}) by $\boldsymbol{m}$ instead of $\boldsymbol{M}$. Then, the success and error probabilities employed in the main text are given by
\begin{eqnarray}
P_s&=&1-P_{(0,0,1,1,0,0,0,0)}-P_{(0,0,0,0,1,1,0,0)}+P_{(0,0,1,1,1,1,0,0)}\nonumber \\
P_{e,1}&=&1-2P_{(0,0,1,1,0,0,0,0)}+P_{(1,1,1,1,0,0,0,0)}\nonumber \\
P_{e,2}&=&1-2P_{(0,0,0,0,1,1,0,0)}+P_{(0,0,0,0,1,1,1,1)}.
\label{append:pspe}
\end{eqnarray}
The exact analytical expressions of those probabilities obtain very extensive forms.

\begin{figure}[t]
\centerline {\includegraphics[width=0.9\linewidth]{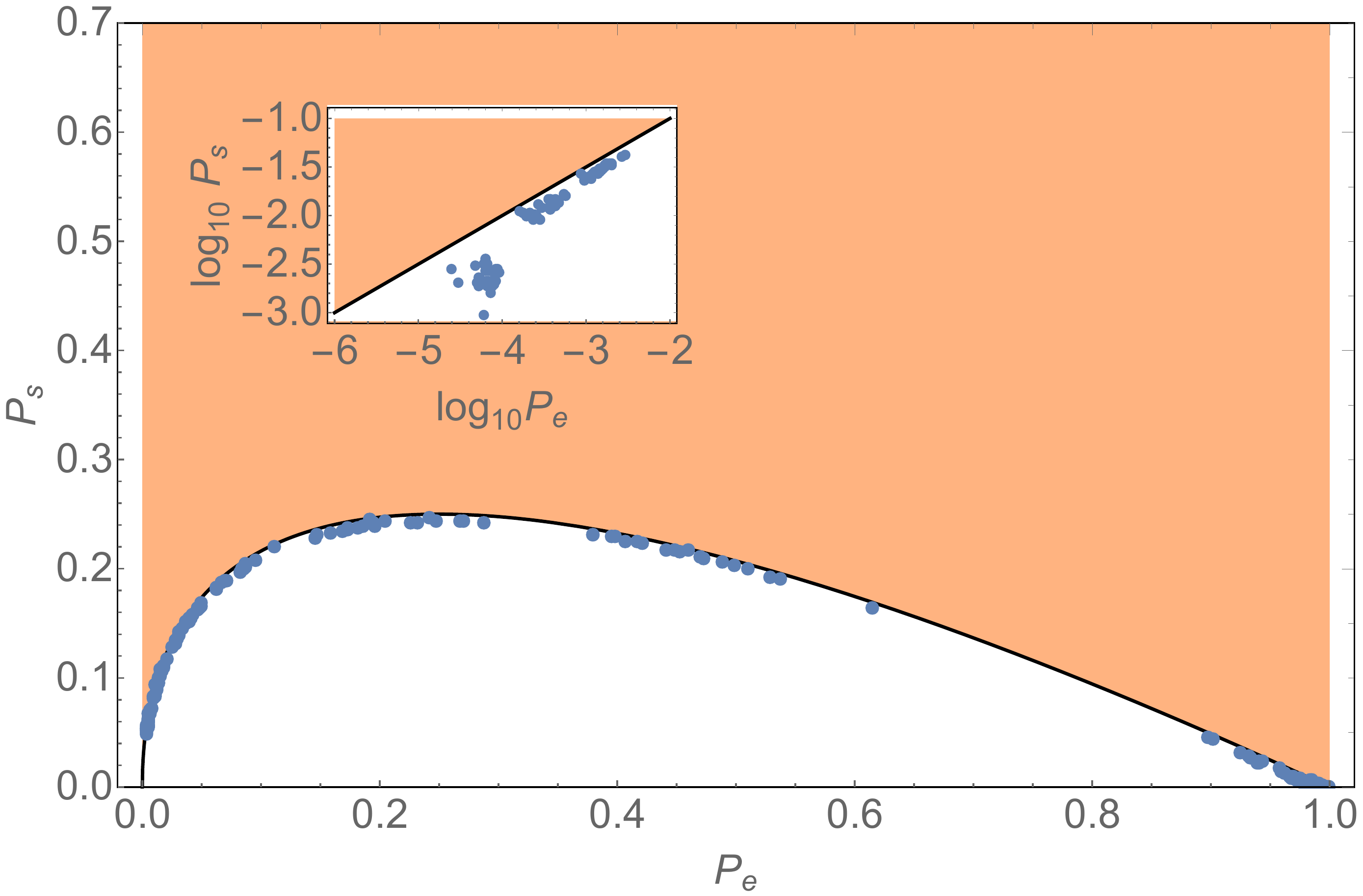}}
\caption{Figure depicts results of a Monte - Carlo simulation randomly producing probabilities $P_{1,1}$, $P_{e,1}=1-P_{0}^{(1)}-P_{0}^{(2)}$ and $P_{e,2}=1-P_{1}^{(2)}-P_{1}^{(2)}$ exhibited by Gaussian states when a PNRD is used for detection. The inset shows the results of experimentally relevant region of states with very low error probabilities. The black solid line corresponds to the threshold covering all the states. The blue points represent fifty points generated in the Monte-Carlo simulation that get closer to the threshold. A total number of cycles in the simulation was $10^6$ for each parameter of squeezing in a single mode. }
\label{fig:mcPNRD}
\end{figure}

The formulas can be modified for the response of a detector distinguishing a number of arriving photons. Two such detectors measuring different modes allow us to get the probabilities $P_{m n}=\langle m \vert \langle n \vert \rho \vert m \rangle \vert n \rangle$ for $m$ and $n$ up to some number. Let us work out the probability $P_{m,n}$ exhibited by Gaussian states. They are achieved from an overlap of Wigner functions
\begin{eqnarray}
P_{m,n}= 16 \pi^2 \int W_m(x_1,p_1)W_n(x_2,p_2)\times \nonumber \\
W_G(x_1,p_1,x_2,p_2)\mathrm{d}x_1\mathrm{d}x_2\mathrm{d}p_1\mathrm{d}p_2,
\label{WigOverLap}
\end{eqnarray}
where $W_m$ stands for the Wigner function of the Fock state $\vert m \rangle$ and $W_G$ denotes the Wigner function of a Gaussian state. Direct calculation of the integral (\ref{WigOverLap}) for a general Gaussian state in two modes gives rise to very extensive expressions, which are hard to manipulate analytically. Therefore, the solution of the integral is expressed in terms of derivation of the formula (\ref{p0G}) according to the elements of the covariance matrix (\ref{covMat}). Let us introduce an operator
\begin{eqnarray}
    \mathcal{L}_{M,i}=-1-2\partial_{\bar{\Gamma}_{2i-1,2i-1}}\nonumber \\
    -2\partial_{\bar{\Gamma}_{2i,2i}},
\end{eqnarray}
where $\boldsymbol{\bar{\Gamma}}=\boldsymbol{\Gamma}+\boldsymbol{M}$ and $i$ being $1$ or $2$ picks relevant elements of $\boldsymbol{\bar{\Gamma}}$.
The probabilities exposing the quantum non-Gaussian coincidences are expressed as
\begin{eqnarray}
P_{1,1} &=& \mathcal{L}_{\boldsymbol{M},1}\mathcal{L}_{\boldsymbol{M},2}P_{\boldsymbol{M}}(\boldsymbol{\Gamma})\nonumber \\
P_{1}^{(i)} &=& \mathcal{L}_{\boldsymbol{M}^{(i)},i}P_{\boldsymbol{M}^{(i)}}(\boldsymbol{\Gamma})\nonumber \\
P_{0}^{(i)} &=& \mathcal{L}_{\boldsymbol{M}^{(i)},i}P_{\boldsymbol{M}^{(i)}}(\boldsymbol{\Gamma}),
\end{eqnarray}
where $\boldsymbol{M}$ is an identity matrix of rank two, $\boldsymbol{M}^{(1)}$ has elements $M_{i,j}^{(1)} =\delta_{i,j} m_i$ with $m_{1}=m_2=1$ and $m_3=m_4=0$ and, finally, $\boldsymbol{M}^{(2)}$ has elements $M_{i,j}^{(2)} =\delta_{i,j} m_i$ with $m_{1}=m_2=0$ and $m_3=m_4=1$.

\section{Derivation of the criteria}
The nonclassicality and the quantum non-Gaussianity reject states that can be prepared as a statistical mixture of coherent and Gaussian states respectively. To prove a density matrix possesses one or both these quantum aspects, we introduce a probability $P_s$ of success and a probability $P_e$ of error and define their linear form
\begin{equation}
    F_a(\rho)=P_s+a P_e
    \label{SMFa}
\end{equation}
with $a$ being a free parameter. The choice of the success and error probabilities can be made arbitrarily for any detection. In this paper, we consider the single-photon avalanche photo-diodes (SPADs) or the photon-number resolving detectors (PNRD) perform a detection.
The main text describes the detection schemes and introduces the probabilities $P_s$ and $P_e$ for both cases.
The criteria stems from optimizing (\ref{SMFa}) over all rejected states giving the threshold function $F(a)$ defined as
\begin{equation}
    F(a):=\max_{\rho \in R} F_a(\rho),
    \label{formalOpt}
\end{equation}
where $R$ generally represents any convex set of states being rejected. Specifically, this paper deals with $R$ being the set of all classical states or the set of all mixtures of the Gaussian states according to the quantum aspect that is examined. Importantly, because $F_a(\rho)$ is linear in the density matrix, the optimum in (\ref{formalOpt}) is achieved by pure states \cite{Filip2011}, which can be always parametrized \cite{Braunstein2005}. With knowledge of the threshold function $F(a)$, both quantum aspects are expressed formally as
\begin{equation}
    \exists a: F_a(\rho)>F(a),
    \label{SM:formalCond}
\end{equation}
where $\rho$ is an inspected state, which can exhibit the quantum aspects.
From the mathematical point of view, the whole procedure can be understood as optimizing $P_s$ with a constraint on the probability $P_e$ as shown further. The criterion can be reformulated according to
\begin{equation}
     \exists a: F_a(\rho)>F(a) \Leftrightarrow P_s>\min_a \left[F(a)-a P_e \right],
     \label{SM:TwoInt}
\end{equation}
where the probabilities $P_s$ and $P_e$ in (\ref{SM:TwoInt}) stand for the success and error probabilities of an inspected state. Further, let $\rho_a \in R$ denotes the state optimizing $F_a(\rho)$ over $\rho \in R$ for a particular parameter $a$. When the global minimum in (\ref{SM:TwoInt})  obeys $\frac{\mathrm{d}}{\mathrm{d} a} \left[ F(a)-a P_e\right]=0$, it can be simply proved that the derivation $\frac{\mathrm{d}}{\mathrm{d} a} F(a)$ equals to to the error probability of the state $\rho_a$. This guarantees the identity of error probabilities between the state $\rho_a$ and the inspected state $\rho$, i. e. it defines a constraint in this optimizing task. We obtain the same identity even when the minimum of $F(a)-a P_e$ occurs in a point where that function is not smooth. 
Thus, the optimizing is equivalent to the Lagrange optimizing task with $a$ being the Lagrange multiplier.

Further, we will focus on the scheme employing SPADs to illustrate the approach explicitly. The other detection with PNRDs can be dealt analogously to derive the criteria.

\subsection{Nonclassicality}
For the coherent states $\vert \alpha \rangle_1 \vert \beta \rangle_2$, the success and error probabilities in the scheme in Fig.~1 a) of the main text obtain
\begin{equation}
    \begin{aligned}
        P_s &=\left(1-e^{-\vert \alpha \vert^2/2}\right)\left(1-e^{-\vert \beta \vert^2/2}\right)\\ \nonumber
        P_{e,1}+P_{e,2} &=\left(1-e^{-\vert \alpha \vert^2/2}\right)^2+\left(1-e^{-\vert \beta \vert^2/2}\right)^2.
    \end{aligned}
\end{equation}
Optimizing the function $F_a(\vert \alpha \vert^2,\vert \beta \vert^2)$ induces the threshold function $F(a)=0$ for $a\leq -1/2$ and $F(a)=1+2 a$ for $a >-1/2$. The condition $P_s>\min_a \left[F(a)-a P_e\right]$ implies
\begin{equation}
    P_s>\frac{1}{2}(P_{e,1}+P_{e,2}),
\end{equation}
which corresponds to the sufficient condition for the nonclassicality.

\subsection{Quantum non-Gaussian coherences}
\subsubsection{Monte-Carlo simulation}
The threshold covering all the mixtures of Gaussian states is induced from maximizing the function
\begin{equation}
\begin{aligned}
F_a(\vert \xi_1 \vert,\vert \xi_2 \vert, \phi, \tau, \vert \alpha_1 \vert, \vert \alpha_2 \vert,\psi_1,\psi_2)=\nonumber \\
P_s+a(P_{e,1}+P_{e,2}),
\end{aligned}
\label{SM:thF}
\end{equation}
where $P_s$, $P_{e,1}$ and $P_{e,2}$ are success and error probabilities defined in Fig.~1 a) of the main text and the arguments of the function $F_a$ represent parameters giving a general two-mode Gaussian state. The probabilities $P_s$, $P_{e,1}$ and $P_{e,2}$ are expressed from (\ref{p0G}) and (\ref{append:pspe}). The criterion implies from the condition
\begin{equation}
\begin{aligned}
&\ & \exists a: P_s+a(P_{e,1}+P_{e,2})>F(a)\\
&=&\max F_a(\vert \xi_1 \vert,\vert \xi_2 \vert, \phi, \tau, \vert \alpha_1 \vert, \vert \alpha_2 \vert,\psi_1,\psi_2).
\end{aligned}
\label{condA}
\end{equation}
Since the function $F_a$ is linear in a state, the threshold function $F(a)$ covers even all mixtures of Gaussian states \cite{Filip2011}.
The maximal state holds
\begin{equation}
    \nabla F_a=0.
    \label{locExtr}
\end{equation}
We are going to show that the two-mode squeezed state
\begin{equation}
    \vert G_r \rangle = \sqrt{1-r^2}\sum_{n=0}^{\infty}r^n \vert n \rangle \vert n \rangle,
    \label{optSt}
\end{equation}
obeys the condition (\ref{locExtr}), and therefore the state belong to a local maximum at least. The state (\ref{optSt}) is induced by the unitary operators with the parameters  $\phi=\pi/2$, $\tau=1/2$ and $\vert \alpha_2 \vert=\vert \alpha_1 \vert=0$ and $ \xi_2=\xi_1$ with $\xi_1$ being real and positive. The function $F_a$ of this state yields
\begin{equation}
    F_a(r)= \frac{r^2 \left[2+r^2+2a r^2(4-r^2)\right]}{(4-r^2)(2-r^2)},
\end{equation}
where $r$ is introduced in (\ref{optSt}) and is given by $r=(1-e^{-2\xi_1})/(1+e^{-2\xi_1})$. The optimum over $r$ happens when $r$ fulfills
\begin{equation}
    a=-\frac{8-4(-2+r^2)r^2}{r(4-r^2)^3}.
    \label{ar}
\end{equation}
Because $r\in (0,1)$, it can be obeyed only for $a\in (-\infty, -4/9)$. Further, let us introduce the operator
\begin{equation}
\begin{aligned}
    D_n(a)=\lim_{t\rightarrow 0}\frac{\mathrm{d^n}}{\mathrm{d}t^n}F_a(\vert \xi_1 \vert,\vert \xi_1 \vert+W t, ...\\
    ...\pi/2+\phi t, 1/2+T t,\vert \alpha_1 \vert t, \vert \alpha_2 \vert t,\psi_1,\psi_2)
\end{aligned}
\end{equation}
and assume $\xi_1$ yields $r$ holding (\ref{ar}). Explicit calculation of the derivatives confirms $D_1(a)=0$ and $D_2(a)<0$ for any $a\in (-\infty, -4/9)$, which satisfies the conditions for the local maximum. If the state (\ref{optSt}) gives the global maximum as well, the requirement (\ref{condA}) will lead to
\begin{equation}
    P_s>\frac{1}{2}\sqrt{\frac{P_e}{8+P_e}}\left[2+P_e+\sqrt{P_e(8+P_e)} \right].
    \label{probCond}
\end{equation}
To certify this, we performed several Monte-Carlo simulations where random Gaussian states $\vert G \rangle_{1,2}$ were generated. Each simulation was performed for a fixed parameter $\xi_1$ and all the other parameters were randomly produced. Changing $\xi_1$ shifted a region of probabilities were a Monte-Carlo simulation set the generated points. We carried out ten simulations with fixed squeezing $\exp(- \vert \xi_1 \vert)=\lbrace 0.1,0.2,0.3,0.4,0.5,0.6,0.7,0.8,0.9 \rbrace$ for both detection schemes. The others parameters were generated randomly in intervals $\exp(- \vert \xi_2 \vert) \in (0,1)$, $\phi \in (0,2\pi)$, $\tau \in (0,1)$, $\vert \alpha_1 \vert \in (0,1.5)$, $\vert \alpha_2 \vert \in (0,1.5)$, $\psi_1 \in (0,2\pi)$ and $\psi_2 \in (0,2\pi)$. The same we did for proving the threshold for quantum non-Gaussian coincidences employing PNRDs. Fig. 2 in the main text presents the results for the measurement with SPADs and Fig.~\ref{fig:mcPNRD} shows the results when PNRDs are used. Since each simulation produced $10^6$ states, the figures show only the best $50$ attempts in each simulation. Because the simulations were carried out ten times with different $\xi_1$, each figure presents five hundred best attempts.

\subsubsection{Approximate solution}
The final condition (\ref{probCond}) follows from a conjecture that the state (\ref{optSt}) represent an optimal state globally. Beside the performed Monte-Simulation, this result can be verified with a certain degree of accuracy from the Taylor series of the success and error probabilities. Let us reinterpret the optimizing of the function (\ref{SM:thF}) as an optimizing of the probability $P_s$ with a constraint $P_e=(P_{e,1}+P_{e,2})/2$ and with $a$ being the Lagrange multiplier. Since the optimal state has to be a pure state, we can determine the optimal state by solving the Lagrangian task for pure Gaussian states in the regime of states with surpassed $P_e$. We will find the solution through five theorems that are based on the following postulate.
\newline

\textbf{Postulate}. Let $\mathcal{T}_{G}$ denotes a class of Gaussian states whose parameters from the Bloch-Messiah reduction are given by the polynomials
\begin{equation}
    \begin{aligned}
        \vert \xi_i \vert &=\sum_{n=1}^{\infty}V_{i,n}t^n\\
        \vert \alpha_i \vert^2 &= \sum_{n=1}^{\infty}A_{i,n}t^n\\
        \tau&=\sum_{n=0}^{\infty} \tau_n t^n \\
        \phi&=\sum_{n=0}^{\infty} \phi_n t^n.
    \end{aligned}
    \label{energyExp}
\end{equation}
with $i=1,2$. Whereas the coefficients $V_{i,n}$, $A_{i,n}$ $\tau_n$ and $\phi_n$ are considered to be fixed for a given class $\mathcal{T}_{G}$, the parameter $t$ can gain arbitrary non-negative value. Then, the state $\rho \in \mathcal{T}_{G}$ exhibits Taylor expansion of its success an error probabilities
\begin{equation}
    \begin{aligned}
        P_s=\sum_{n=2}^{\infty}S_n t^n\\
        P_e=\sum_{n=2}^{\infty}E_n t^n,
    \end{aligned}
    \label{ThTS}
\end{equation}
where $S_n$ and $E_n$ depend on the set of parameters $\left\{V_{i,1},...,V_{i,n-1}, A_{i,1},...,A_{i,n-1}, \tau_{0},...,\tau_{n-2}, \phi_{0},...,\phi_{n-2}\right\}$. Note, $\mathcal{T}_{G}$ and $t$ do not generally specify unambiguously the parameters $S_n$ and $E_n$ since $S_n$ and $E_n$ also depend on the angles $\psi_1$ and $\psi_2$ when $\alpha_{1,2} \neq 0$.  The task here is to identify the set of Gaussian states $\mathcal{T}_{G}$ together with $\psi_1$ and $\psi_2$ (if $\alpha_{1,2} \neq 0$) that represent solution of the Lagrange optimizing task (\ref{formalOpt}). A particular choice of $a$ in (\ref{formalOpt}) differentiates the optimal Gaussian states in $\mathcal{T}_{G}$ only by the parameter $t$.
\newline

\textbf{Theorem 1}. The Gaussian states can satisfy $E_2=0$ and $E_3=0$ in the Taylor series if and only if (\emph a) $V_{1,1}=V_{2,1}$, $A_{1,1}=A_{2,1}=0$, $\tau_0=1/2$ and $\phi_0=\pi/2$ or (\emph b) $V_{1,1}=V_{2,1}=0$, $A_{1,1}=A_{2,1}=0$.

\emph{Proof}. We first determine Gaussian states that exhibit $E_2=0$ in the Taylor series. According to Postulate, it suffices to consider $\vert \xi_i \vert=V_{i,1} t$, $\vert \alpha_i \vert^2=A_{i,1}t$, $\tau=\tau_0$, $\phi=\phi_0$ and make the Taylor expansion of $P_e$ with respect to $t$. It works out to be
\begin{equation}
\begin{aligned}
&\ E_2 =\frac{1}{64}\left\{A_{1,1}^2+A_{2,1}^2+4\left[1-2(1-\tau_0)\tau_0\right](V_{1,1}^2+V_{2,1}^2)\right.\\ \nonumber
&-4 A_{1,1}\tau_0 V_{1,1} \cos 2\psi_1+4 A_{1,1} (1-\tau_0) V_{2,1} \cos 2(\psi_1+\phi_0)\\ \nonumber
&-4(1-\tau_0) A_{2,1} \tau_0 V_{1,1} \cos 2\psi_1+4 \tau_0 A_{2,1} V_{2,1} \cos 2(\psi_1+\phi_0) \\ \nonumber
&\left.+8(1+\tau_0)\tau_0 V_{1,1}V_{2,1}\cos 2\phi \right\} \geq 0,
\end{aligned}
\end{equation}
where the relation in the end of the expression implies from the requirement that the probability $P_e$ is not negative. It holds for all the physically well defined parameters $V_{i,1}>0$ and $\tau\in (0,1)$. 
Let us further define the quadratic function $\widetilde{E}_{V_{1,1},\phi,\psi_1,\psi_2}$
\begin{equation}
\begin{aligned}
    \widetilde{E}_{V_{1,1},\phi_0,\psi_1,\psi_2}(A_{1,1},A_{2,1},\tau_0,V_{2,1}) \nonumber \\
    \equiv E_2(V_{1,1},\phi_0,\psi_1,\psi_2,A_{1,1},A_{2,1},\tau_0,V_{2,1}),
\end{aligned}
\end{equation}
where $V_{1,1},\phi,\psi_1,\psi_2$ represent parameters of the function. The optimal variables $A_{1,1}$,$A_{2,1}$, $V_{2,1}$ and $\tau$ fulfill
\begin{equation}
    \nabla \widetilde{E}_{V_{1,1},\phi_0,\psi_1,\psi_2}(A_{1,1},A_{2,1},\tau_0,V_{2,1})=0.
    \label{minE}
\end{equation}
The roots are given by solving a set of four linear equations determining when $\widetilde{E}_{V_{1,1},\phi_0,\psi_1,\psi_2}=0$, and therefore it allows us to identify all the parameters giving $E_{2}=0$.

The next step is identification of all parameters that satisfy (\ref{minE}) and $E_3=0$. According to Postulate, we have to consider $\vert \xi_i \vert=V_{i,1} t+V_{i,2}t^2$, $\vert \alpha_i \vert^2=A_{i,1}t+A_{i,2}t^2$, $\tau=\tau_0+\tau_1 t$ and $\phi=\phi_0+\phi_1 t$. Inserting that into the formulas (\ref{append:pspe}) and expanding it with respect to $t$ leads formally to
\begin{equation}
\begin{aligned}
    E_3&=f_0+f_{V,1} V_{1,2}+f_{V,2} V_{2,2}\\
    &+f_{A,1}A_{1,2}+f_{A,2}A_{2,2}+f_{\tau}\tau_1+f_{\phi}\phi_1,
    \end{aligned}
\end{equation}
where $f_0$, $f_{V,1}$, $f_{V,2}$, $f_{A,1}$, $f_{A,2}$, $f_{\tau}$, $f_{\phi}$ are some functions independent of $V_{i,2}$, $A_{i,2}$, $\tau_1$ and $\phi_1$. By direct substitution, we can verify that (\ref{minE}) implies $f_{V,1}=f_{V,2}=f_{A,1}=f_{A,2}=f_{\tau}=f_{\phi}=0$ for any $\tau_0 \in (0,1)$ and $\phi_0$, and therefore $E_3$ becomes independent of $V_{1,2}$, $V_{2,2}$, $A_{1,2}$, $A_{2,2}$, $\tau_1$ and $\phi_1$. Further, we checked that $f_0=0$ and (\ref{minE}) are satisfied if and only if $\phi_0=\pi/2$ or $V_{1,1}=0$. Using (\ref{minE}) equations for $\phi_0=\pi/2$ gives rise to $V_{2,1}=V_{1,1}$, $\tau_0=1/2$ and $A_{1,1}=A_{2,1}=0$. Contrary, $V_{1,1}=0$ leads to a trivial solution $V_{2,1}=A_{1,1}=A_{2,1}=0$. \qedsymbol
\newline

\textbf{Theorem 2}. The optimal Gaussian states fulfill $E_2=E_3=0$.

\emph{Proof}. According to Theorem 1, there exists a set of Gaussian states $\widetilde{\mathcal{T}}_{G}$ exhibiting $E_2=E_3=0$. We show that any set of Gaussian states $\mathcal{T}'_G$ giving $E_3>0$ can not represent the optimal states. Note, the coefficients $E_2$ and $E_3$ also depends on the angles $\psi_{1,2}$ for a given set $\mathcal{T}'_G$. However, this dependence is not important in this proof.

Assuming the set $\mathcal{T}'_G$ includes the optimal states only, we can chose $t$ being so small that the success and error probabilities become $P_s \approx S_{2} t^2$ and $P_e \approx E_{3} t^3$. Then, the Gaussian states that belong to any $\mathcal{T}_G$ should obey
\begin{equation}
    P_s^3 \leq \frac{S_2^3}{E_3^2}P_e^{2}
    \label{SM:AppPr}
\end{equation}
for sufficiently small $t$. However, $\rho_{g} \in \widetilde{T}_G$ exhibits in this limit $P_s \approx \widetilde{S}_2 t^2$ and $P_e \approx \widetilde{E}_4 t^4$. It violates the condition (\ref{SM:AppPr}) for $t<\min \left\{\widetilde{S}_{2}^3 E_3^2/S_2^3/\widetilde{E}_{4}^2,\epsilon\right\}$, where $\epsilon \ll 1$, and therefore the states having $E_3>0$ can not be optimal. To prove this for states having $E_2>0$ is analogous. \qedsymbol
\newline

\textbf{Theorem 3}. The parameters $\tau$ and $\phi$ determining the optimal Gaussian states have the Taylor series
\begin{equation}
    \begin{aligned}
        \tau &=\frac{1}{2}+\sum_{n=1}^{\infty} \tau_n t^n \\
        \phi &=\frac{\pi}{2}+\sum_{n=1}^{\infty} \phi_n t^n.
    \end{aligned}
    \label{TSTPhi}
\end{equation}
The proof directly follows from Theorem 1 and Theorem 2. 
\newline

\textbf{Theorem 4}. The Gaussian states yields $E_2=E_3=E_4=0$ if and only if $V_1=V_2=0$, $A_{1,2}=A_{2,2}=0$ and $V_{1,2}=V_{2,2}$. For fixed $V_1>0$, $E_4$ reaches its minimum for $A_{1,2}=A_{2,2}=0$, $\phi_{1}=\tau_1=0$ and $V_{2,1}=V_{2,2}$.

\emph{Proof}. Expanding the expression for $P_e$ with $A_{1,1}=A_{2,1}=0$ and $V_{2,1}=V_{1,1}$, we obtain
\begin{equation}
    \begin{aligned}
        E_4&=\frac{1}{64}\left[ A_{1,2}^2+A_{2,2}^2+4(A_{1,2}+A_{2,2}+2\phi_1^2+8 \tau_1^2)V_{1,1}^2 \right.\\
        &+4 V_{1,1}^4+2(V_{1,2}-V_{2,2})^2\\
        &+2 A_{1,2}(4\tau_1 V_{1,1}+V_{1,2}-V_{2,2})\cos(2\psi_1)\\
        &+2 A_{2,2}(4\tau_1 V_{1,1}-V_{1,2}+V_{2,2})\cos(2\psi_1)\\
        &\left.-4 \phi V_{1,1}(A_{1,2}\sin(2\psi_1)+A_{2,2}\sin(2\psi_2)\right],
    \end{aligned}
\end{equation}
where $\tau_1$ and $\phi_1$ are introduced in (\ref{TSTPhi}). Let us introduce the function
\begin{equation}
\begin{aligned}
    &\ \widetilde{E}_{V_{1,1},\phi_1,\psi_1,\psi_2}(V_{1,2},V_{2,2},\tau_1,A_{1,2},A_{2,2},\psi_1,\psi_2) \\
    &\equiv E_4(V_{1,1},V_{1,2},V_{2,2},\phi_1,\tau_1,A_{1,2},A_{2,2},\psi_1,\psi_2).
\end{aligned}
\end{equation}
The identity $\nabla \widetilde{E}_{V_{1,1},\phi_1,\psi_1,\psi_2}=0$ is obeyed only if
\begin{equation}
\begin{aligned}
   \phi_1&=V_{1,1}(\mbox{ctan}\psi_2+2\mbox{ctan}^2 2 \psi_1 \sin 2 \psi_2+\tan \psi_2)\\
  &\times \sin 2\psi_1  /(\sin 2\psi_1 +\sin 2\psi_2)\\
  A_{1,2}&=2 V_{1,1}\frac{\phi_1\sin 2 \psi_1 -V_{1,1}}{\sin^2 \psi_1}\\
  A_{2,2}&=2 V_{1,1}\frac{\phi_1\sin 2 \psi_2 -V_{1,1}}{\sin^2 \psi_2}.
\end{aligned}
\end{equation}
It can be verified that no angles $\psi_1$ and $\psi_2$ satisfy the equations for $V_{1,1}>0$ and $A_{i,2}\geq 0$. The latter constraint implies from (\ref{energyExp}) where $A_{i,2}$ are the first non-zero coefficients. It means $\nabla \widetilde{E}_{V_{1,1},\phi_1,\psi_1,\psi_2}=0$ does not identify the optimum. Thus, we set $A_{2,2}=0$. After manipulation with the equations, we arrive at
\begin{equation}
   E_4=\frac{1}{512}(A_{1,2}^2+16 A_{1,2}V_{1,1}^2+16 V_{1,1}^4-A_{1,2}^2 \cos 4 \psi_1).
\end{equation}
It gains its minimum $E_4=V_{1,1}^4 /32$ for $A_{1,2} \geq 0$ when $A_{1,2}=0$. Consequently, it induces $\tau_1=\phi_1=0$ and $V_{1,2}=V_{2,2}$. \qedsymbol

Theorems 1-3 allows us to make the most rough approximation of the threshold for quantum non-Gaussianity. According to them, we get that the optimal states have to exhibit 
\begin{equation}
    \begin{aligned}
        P_s=\frac{1}{16}t^2+\sum_{n=3}^{\infty}S_n t^n\\
        P_e=\frac{1}{32}t^4+\sum_{n=5}^{\infty}E_n t^n.
    \end{aligned}
    \label{RApp}
\end{equation}
This approach is extendable for determining the higher coefficients in the Taylor expansion. To do that, we prove the following.
\newline

\textbf{Theorem 5}. If the parameters of the optimal Gaussian states have the Taylor series (\ref{TSTPhi}), the criterion of quantum non-Gaussianity obtains a form
\begin{equation}
    P_s>\sum_{n=1}^{\infty}T_n \left(\sqrt[4]{P_e}\right)^n,
    \label{TnTS}
\end{equation}
where $T_n$ is some function of parameters $\lbrace E_4,..,E_{n+3},S_2,...,S_{n+1} \rbrace$ introduced in Postulate.

\emph{Proof}. We can conclude immediately from relations (\ref{RApp}) that the Taylor series is some summation of members $\sqrt[4]{P_e}^n$ with $n\geq 2$. We will provide a procedure giving the coefficients $T_n$. Let us define a function 
\begin{equation}
    f(t)=\sqrt[4]{\sum_{n=4}^{\infty}E_n t^n}.
\end{equation}
The inverse function $g(f) \equiv f^{-1}(t)$ has a Taylor series
\begin{equation}
    g=\sum_{n=1}^{\infty} G_n f^n,
    \label{gTS}
\end{equation}
where $G_n=\lim_{f\rightarrow 0}\frac{\mathrm{d} g(f)}{\mathrm{d} f}/n!$. These derivations can be achieved from deriving $n$ times both sides of the identity
\begin{equation}
    g\circ f(t)=t.
\end{equation}
From the first derivation, we get $\frac{\mathrm{d}g}{\mathrm{d}f}=1/(\frac{\mathrm{d}f}{\mathrm{d}t})$. Deriving it two times leads to $\frac{\mathrm{d^2}g}{\mathrm{d}f^2}=-\frac{\mathrm{d}^2f}{\mathrm{d}t^2}/(\frac{\mathrm{d}f}{\mathrm{d}t})^3$ etc. Consequently, the $n$th derivation $\frac{\mathrm{d}^n g}{\mathrm{d}f^n}$ depends only on $\left \{\frac{\mathrm{d}f}{\mathrm{d}t},..., \frac{\mathrm{d}^{n-1}f}{\mathrm{d}t^{n-1}}\right \}$. 
Because the inverse function $g$ returns the parameter $t$ as a function of $\sqrt[4]{P_e}$ according to the identity $P_e=\sum_{n=4}^{\infty}E_n t^n$,
substituting $t$ in $\sum_{n=2}^{\infty} S_n t^n$ by $g$ from (\ref{gTS}) determines the dependence of $T_n$ on the parameters $\lbrace E_4,..,E_{n+3},S_2,...,S_{n+1} \rbrace$. \qedsymbol
\newline

The proof of Theorem 5 instructs us how to find coefficients $T_n$ in (\ref{TnTS}). Their sequential optimizing over the parameters from (\ref{energyExp}) enables derivation of the Taylor series of the threshold function exposing the quantum non-Gaussian coincidences. From Theorems 1-3, we get immediately $T_1=\frac{1}{2\sqrt{2}}$. Let us determine $T_2$ and $T_3$ to illustrate the procedure. According to the proof of Theorem 5, we obtain
\begin{equation}
    T_3=\frac{2E_4 S_3-E_5 S_2}{2 E_4^{7/4}}=-\frac{A_{1,3}+A_{2,3}}{4 V_{1,1}^3},
    \label{T3}
\end{equation}
which holds when $E_2=E_3=0$ and $T_1=\frac{1}{2\sqrt{2}}$. Since $A_{1,3}$ and $A_{2,3}$ are first member of Taylor series giving non-negative value, they are also non-negative, and therefore (\ref{T3}) is optimal for $A_{1,3}=A_{2,3}=0$ giving $T_3=0$. The following member obtains
\begin{equation}
    \begin{aligned}
        T_4 &=\frac{2 E_5^2 S_2-2E_4 E_6 S_2-3 E_4 E_5 S_3+4 E_4^2 S_4}{4 E_4^3}\\
        &=\frac{-(V_{1,3}-V_{2,3})^2-2(A_{1,4}+A_{2,4}+2 \phi_2^2+8 T_2^2)V_{1,1}^2}{4 V_{1,1}^6}\\
        & + \frac{1}{4},
    \end{aligned}
\end{equation}
which acquires its optimum $T_4=1/4$ from the same reasons.


\begin{figure*}[ht!]
\centerline {\includegraphics[width=0.9\linewidth]{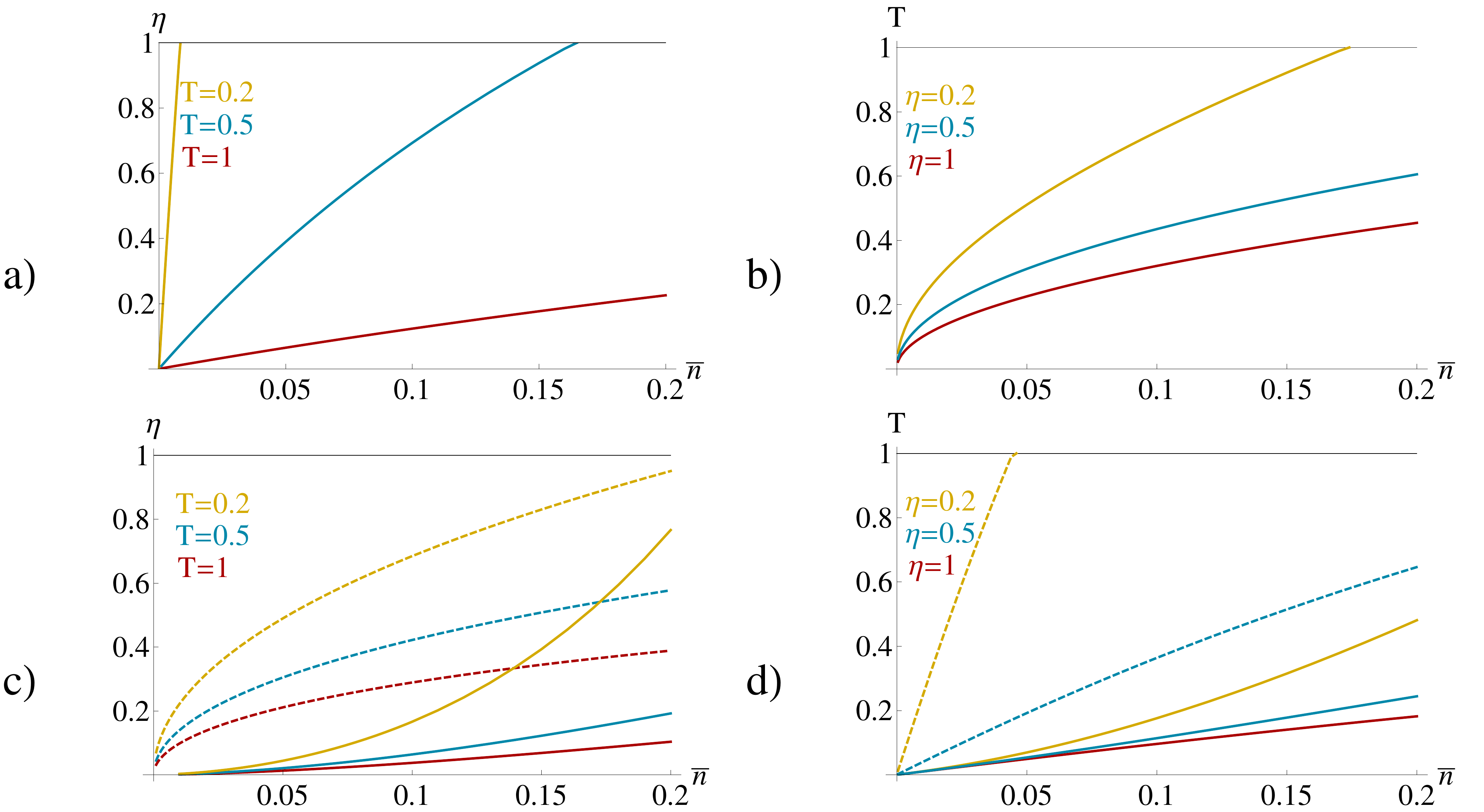}}
\caption{Figures present quantum non-Gaussian coincidences and quantum non-Gaussianity in different detection scenarios. The considered model is a correlated photon pair $\eta \vert 1 \rangle \langle 1 \vert\otimes \vert 1 \rangle \langle 1 \vert+(1-\eta) \vert 0 \rangle \langle 0 \vert\otimes \vert 0 \rangle \langle 0 \vert$ that is deteriorated by indistinguishable Poissonian noise with the mean number of photons $\bar{n}$ and losses. a) Thresholds revealing the quantum non-Gaussian coincidences of the state for different losses $T$, which the colors distinguish. Whereas the solid lines correspond to conditions when SPADs are employed, the dashed lines represent the conditions for measurement with PNRDs. The quantum non-Gaussian states are above those lines. b) The solid lines represent robustness of the quantum non-Gaussian coincidences against losses for several parameters $\eta$ using detection with SPADs. The dashed lines show the same thresholds when PNRDs are used. c) Employing the criterion in \cite{Lachman2013}, the figure presents threshold parameters for the quantum non-Gaussianity of the heralded state (solid) and unheralded state (dashed). The colors differentiate losses $T$ again. d) An analysis regarding the robustness against losses of the quantum non-Gaussian test for the heralded state (solid) and unheralded state (dashed). The colors distinguish probability of the photon-pair emission $\eta$. The robustness of both heralded and unheralded states is identical for $\eta=1$.}
\label{fig:model}
\end{figure*}

\subsection{Single-mode quantum non-Gaussianity}
The single-mode quantum non-Gaussianity manifests itself in a detection scheme where a beam-splitter divides the light between two SPADs as Fig.~3 of the main text depicts. The response of such a detector on the pure Gaussian states is determined from the no-click probabilities
\begin{equation}
\begin{aligned}
    P_{(1,1)}&=2\sqrt{V}\frac{e^{-\frac{|\alpha|^2 \left[(1+V)+(1-V)\cos 2 \phi \right]}{4(1+V)}}}{1+V},\\
    P_{(1,0)}&=2\sqrt{\frac{V}{3 V^2+10 V+3}}\frac{e^{-\frac{|\alpha|^2 \left[1+6V+V^2+(1-V^2)\cos 2 \phi \right]}{4(1+3V)(3+V)}}}{1+V}\\
    P_{(0,1)}&=P_{(1,0)}
\end{aligned}
\end{equation}
where $P_{\boldsymbol{M}}$ with the vector $\boldsymbol{M}$ was defined in Section I and the parameters $|\alpha|$ and $V=\exp(-2|\xi|)$ identify the displacement and squeezing operators that generate the Gaussian states. The criterion incorporates the probability $P_s$ of success and the probability $P_e$ of error that are given by
\begin{equation}
\begin{aligned}
    P_s&=1-P_{(1,0)},\\
    P_e&=1-2P_{(1,0)}+P_{(1,1)}.
\end{aligned}
\end{equation}
The threshold function $F(a)$ stems from maximizing the combination of probabilities $F_a(V,|\alpha|,\phi)=P_s+a P_e$ over the parameters $V$, $|\alpha|$ and $\phi$. The equation $\nabla F_a=0$ is satisfied by the parameters holding $\phi=0$ and $|\alpha |^2=(3+V-3V^2-V^3)/V/(1+3V)$, which eliminates the angle $\phi$ and the amplitude $|\alpha|$. The dependence of the remaining parameter $V$ on the choice of the parameter $a$ can not be expressed analytically. However, interpreting the task as Lagrange optimizing task allows us to exclude $a$ and express the threshold for single-mode quantum non-Gaussianity according to \cite{Lachman2013}
\begin{equation}
\begin{aligned}
    P_s&=1-4e^{-\frac{1-V^2}{2V(1+3V)}}\sqrt{\frac{V}{3+10 V+3 V^2}},\\
    P_e&=1-8e^{-\frac{1-V^2}{2V(1+3V)}}\sqrt{\frac{V}{3+10 V+3 V^2}}\\
    &+2e^{-\frac{3-2V-V^2}{2V(1+3V)}}\frac{\sqrt{V}}{1+V}.
\end{aligned}
\end{equation}
It exposes the single-mode quantum non-Gaussianity when the pair of probabilities $(P_s,P_e)$ surpasses that threshold.

\section{Multi-mode states}
The condition (\ref{probCond}) is applicable only on states occupying two modes. Further, we derive an experimentally relevant condition on multi-mode states determining when those states surpass a threshold covering all the state of the form 
\begin{equation}
    \vert G_N \rangle=\Pi_{i=1}^N \otimes \vert \lambda_i \rangle,
    \label{NModesModel}
\end{equation}
where $\vert \lambda_i \rangle =\sqrt{1-\lambda_i}\sum_{n=0}^{\infty}\left( \sqrt{\lambda_i}\right)^n \vert n \rangle \vert n \rangle$ is the two-mode squeezed state. Let us note, such threshold does not cover provably all the Gaussian states occupying $2N$-modes but it only excludes all  considered states (\ref{NModesModel}) when it is surpassed. The condition is derived for the scheme in Fig.~1 a) of the main text. Again, we focus on the region of states with surpassed error probabilities. Then, the success and error probabilities of the states (\ref{NModesModel}) are expanded according to
\begin{equation}
    \begin{aligned}
    P_s & =\sum_{n=1}^{\infty}S_n t^n\\
    P_e & =\sum_{n=2}^{\infty} E_n t^n.
    \end{aligned}
\end{equation}
where $S_n$ and $E_n$ are some coefficients. According to Theorem 5 the threshold takes the Taylor series
\begin{equation}
    P_s=\sum_{n=1}^{\infty}T_n \left(\sqrt{P_e}\right)^n,
\end{equation}
where $T_n$ are some functions of parameters $\left \{ S_1,...,S_n,E_2,...,E_{n+1} \right \}$. The proof of Theorem 5 provide us with an approach identifying the dependence. We get explicitly the first three members
\begin{equation}
    \begin{aligned}
        T_1 & =\frac{S_1}{\sqrt{E_2}}\\
        T_2 &=\frac{2 S_2 E_2-E_3 S_1}{2 E_2^2}\\
        T_3 &=\frac{5 E_3^2 S_1-4 E_2 E_4 S_1-8 E_2 E_3 S_2+8E_2^2 S_3}{8E_2^{7/2}}.
    \end{aligned}
    \label{Tns}
\end{equation}
The success and error probabilities exhibited by the state $\Pi_{i=1}^N \otimes \vert \lambda_i \rangle$ read
\begin{equation}
    \begin{aligned}
        P_s&=1- 2\Pi_{i=1}^N \frac{1}{1+\lambda_i/2}+4^N\Pi_{i=1}^N \frac{1}{4+3\lambda_i}\\
        P_e&=1- 2\Pi_{i=1}^N \frac{1}{1+\lambda_i/2}+\Pi_{i=1}^N \frac{1}{1+\lambda_i}
    \end{aligned}
    \label{modelPsPe}
\end{equation}
Further, we express $\lambda_i$ as a polynomial
\begin{equation}
    \lambda_i=a_i t+b_i t^2+c_i t^3,
    \label{pols}
\end{equation}
where $t$ is a parameter. Putting the polynomials (\ref{pols}) into (\ref{modelPsPe}) and expanding it with respect to $t$ results in explicit dependence of coefficient $T_n$ in (\ref{Tns}) on $a_i$, $b_i$ and $c_i$, where $i\in (1,...,N)$. Namely, $T_1$ works out to be
\begin{equation}
    T_1=\frac{ N \bar{a}}{2\sqrt{N \overline{a^2}+N^2 \overline{a}^2}},
\end{equation}
where $\bar{a}=\sum_{i=1}^N a_i/N$ and $\overline{a^2}=\sum_{i=1}^N a_i^2/N$. The optimum of $T_1$ is given by $\nabla T_1=0$ and it induces
\begin{equation}
    a_j \overline{a_{\left[j\right]}} =\overline{a^2_{\left[ j\right]}}
    \label{aj}
\end{equation}
for every $j$, where $\overline{a_M}=\sum_{i \not\in M}a_i$ and $\overline{a^2_M}=\sum_{i \not\in M}a_i^2$. Substitution of $a_j$ from (\ref{aj}) in $ a_k\overline{a_{\left[k\right]}} =\overline{a^2_{\left[k\right]}}$ with $k \neq j$ yields
\begin{equation}
   a_k \overline{a_{\left[j,k\right]}} =\overline{a^2_{\left[ j,k\right]}}.
\end{equation}
This operation preserves the equations but reduces their number. Carrying out this operation $N-2$ times, we arrive at two last equations $a_m a_n=a_n^2$ and  $a_n a_m=a_m^2$ with $m\neq n$ having a solution $a_m=a_n$, which directly implies that all $a_i$ are identical and the optimal $T_1$ becomes
\begin{equation}
    T_1=\frac{N}{2\sqrt{ N(N+1)}}
\end{equation}
Let us denote the common coefficient by $a$ and use $\lambda_i=a t+b_i t^2+c_i t^3$ to expand the success and error probabilities according to $t$ again. 
It allows us to obtain $T_2=(5+3 n)/8/(n+1)$, which is independent of parameters $b_i$. The following coefficient works out to be
\begin{equation}
    T_3=\frac{48(\overline{b}^2-\overline{b^2})+a^4(1+N)(2+N)(4+3N)}{48 a^4 (1+N)^2\sqrt{N(1+N)}},
\end{equation}
where $\overline{b}$ and $\overline{b^2}$ are defined analogously to $\overline{a}$ and $\overline{a^2}$. Since we have
\begin{equation}
    \overline{b}^2-\overline{b^2}=-\frac{1}{N^2}\sum_{i=1}^N\sum_{j=1}^N(b_i-b_j)^2,
\end{equation}
the optimum happens when $b_i=b_j$ for all pairs $b_i$ and $b_j$. Finally, we get the optimum
\begin{equation}
    T_3=\frac{(2+N)(4+3N)}{48(1+N)\sqrt{N(1+N)}}.
\end{equation}
If $N$ goes to infinity, we arrive at the condition
\begin{equation}
    P_s>\frac{1}{2}\sqrt{P_e}+\frac{3}{8}P_e+\frac{1}{16}P_e^{3/2}.
\end{equation}
Note, this condition is determined from probabilities of a state $\vert \lambda \rangle^{\otimes N}$. However, this state does not determine the threshold probabilities generally. The following members of the Taylor series are not given by parameters $\lambda_i$ identical in all modes.

\section{Model of realistic states}
A model of a consider realistic source producing a photon pair has the form 
\begin{equation}
\begin{aligned}
     \rho &=\mbox{Tr}_{3,4}\{ L_{2,4}(T_2)L_{1,3}(T_1) \cdot \left[\mathcal{N}_{\bar{n}_1,\bar{n}_2}(\rho_{1,2}) \otimes \vert 0 \rangle_3 \langle 0 \vert \right. \\
    &\left. \otimes \vert 0 \rangle_4 \langle 0 \vert \right] \cdot L_{2,4}^{\dagger}(T_2)L_{1,3}^{\dagger}(T_1)\},
\end{aligned}
\label{modelSM}
\end{equation}
where $\rho_{1,2}(\eta)=\eta \vert 1 \rangle_1 \langle 1 \vert\otimes \vert 1 \rangle_2 \langle 1 \vert+(1-\eta) \vert 0 \rangle_1 \langle 0 \vert\otimes \vert 0 \rangle_2 \langle 0 \vert$, $L_{i,j}$ denotes the unitary operator describing losses from $i$th mode to the $j$th mode, which is traced over, and $\mathcal{N}_{\bar{n}_1,\bar{n}_2}$ is a trace preserving map defined as
\begin{equation}
\begin{aligned}
    &\mathcal{N}_{\bar{n}_1,\bar{n}_2}(\rho)=\frac{1}{4\pi^2}\int \mathrm{d}\phi_1 \mathrm{d}\phi_2 D_1\left[\sqrt{\bar{n}_1}\exp(i \phi_1)\right]\times\\
    &D_2\left[\sqrt{\bar{n}_2}\exp(i\phi)\right] \rho D^{\dagger}_2\left[\sqrt{\bar{n}_2}\exp(i\phi)\right] \times\\
    &D^{\dagger}_1\left[\sqrt{\bar{n}_1}\exp(i \phi_1)\right]
\end{aligned}
\end{equation}
with $D_i(\alpha)$ being the displacement operator acting on the $i$th mode.
Thus, the map $\mathcal{N}_{\bar{n}_1,\bar{n}_2}$ represents effects of noise deteriorating the state. The state (\ref{modelSM}) yields the no-click probabilities
\begin{eqnarray}
P_{(0,0,1,1,1,1,0,0)}&=&\left[1-\eta +\eta \left(1-\frac{T}{2} +\frac{\bar{n}T^2}{4}\right)^2 \right]e^{-T\bar{n}/2}\nonumber \\
P_{(0,0,1,1,0,0,0,0)}&=&P_{(0,0,0,0,1,1,0,0)}\nonumber \\
&=&\frac{1}{4}\left[4+\eta T(-2+\bar{n}T\right]e^{-T\bar{n}/2}\nonumber \\
P_{(1,1,1,1,0,0,0,0)}&=&P_{(0,0,0,0,1,1,1,1)}\nonumber \\
&=&\left[1-\eta T(1-\bar{n}T)\right]e^{-T\bar{n}},
\end{eqnarray}
which allow us to express the success and error probabilities according to (\ref{append:pspe}). When a photon-number resolving detector responds to the model state, the success and error probabilities become
\begin{eqnarray}
P_s &=& \left[\eta T^2(1+\bar{n}-3\bar{n} T+\bar{n}^2 T^2)^2+(1-\eta)T^2 \bar{n}^2 \right]e^{-2T\bar{n}}\nonumber \\
P_{e,1}&=&P_{e,2}=1-\nonumber \\
&\ &\left[1+\eta \bar{n}^2 T^3+\bar{n}(T-2\eta T^2)\right] e^{-T\bar{n}}.
\end{eqnarray}
A condition imposed on the parameters that is required by the quantum non-Gaussian coincidences is presented in Fig.~\ref{fig:model} a) and b) for cases of employing SPADs or PNRDs.

Other experimental scenarios detect the quantum non-Gaussianity in a single mode, where the other mode is either ignored, or used for heralding. In the former case, the state $\rho_1=\mbox{Tr}_2\left[\rho \right]$ works out to be
\begin{equation}
\begin{aligned}
    \rho_r&=\mbox{Tr}_2\left\{ U_{BS}(\tau) D_1(\sqrt{\bar{n}})\left[\eta\vert 1 \rangle_1 \langle 1 \vert \right. \right.\\
    &\left. \left.+(1-\eta)\vert 0 \rangle_1 \langle 0 \vert D_1^{\dagger}(\sqrt{\bar{n}}) \right] \otimes \vert 0 \rangle_2 \langle 0 \vert \right\}
    \end{aligned}
    \label{sm1}
\end{equation}
When the state is prepared conditionally by heralding, the density matrix obtains the same form with $\eta$ increased according to
\begin{equation}
    \eta \rightarrow \eta T \frac{1-e^{-\bar{n}T}(1-T+\bar{n}T^2)}{1+e^{-\bar{n}T}\left[-1+\eta T(1-\bar{n}T)\right]}.
\end{equation}
 Fig.~\ref{fig:model} c) and d) present manifestation of the quantum non-Gaussianity in these cases.

\end{document}